\documentclass[%
reprint,
amsmath,amssymb,
aps,
prb,
]{revtex4-1}

\usepackage{times}
\usepackage[T1]{fontenc}
\usepackage[ansinew]{inputenc}
\usepackage{graphicx}
\usepackage{epstopdf}
\usepackage{dcolumn}
\usepackage{bm}

\usepackage{hyperref}
\hypersetup{
final,
colorlinks=true,
linkcolor=blue,
citecolor=blue,
filecolor=blue,
urlcolor=blue}

\begin{document}

\title{Flexibility of the quasi-non-uniform exchange-correlation approximation}

\author{H. Levämäki}
\email{hpleva@utu.fi}
\author{M.P.J. Punkkinen}
\author{K. Kokko} 
\affiliation{Department of Physics and Astronomy, University of Turku, FI-20014 Turku, Finland}
\affiliation{Turku University Centre for Materials and Surfaces (MatSurf), Turku, Finland}
\author{L. Vitos}
\email{levente@kth.se}
\affiliation{Applied Materials Physics, Department of Materials
Science and Engineering, Royal Institute of Technology, 
Stockholm SE-100 44, Sweden}
\affiliation{Department of Physics and Astronomy, Division of Materials Theory, Uppsala University, Box 516, SE-75121, Uppsala, Sweden} 
\affiliation{Wigner Research Centre for Physics, Institute for Solid State Physics and Optics, H-1525 Budapest, P.O. Box 49, Hungary}
\date{\today}

\begin{abstract}
In our previous study [Phys. Rev. B \textbf{86}, 201104 (2012)] we introduced the so called quasi-non-uniform gradient-level exchange-correlation approximation (QNA) and demonstrated it's strength in producing highly accurate equilibrium volumes for metals and their alloys within the density-functional theory. In this paper we extend the scheme to include the accuracy of bulk modulus as an additional figure of merit and show that this scheme is flexible enough as to allow the computation of accurate equilibrium volumes and bulk moduli at the same time. The power and feasibility of this scheme is demonstrated on NiAl and FeV binary alloys.
\end{abstract}

\pacs{71.15.Mb, 71.15.Nc, 64.30.Ef, 71.20.Be}

\maketitle

\section{\label{sec:intro}Introduction}
Density-functional theory\cite{HK1964,KS1965} has become the most widely used method in determining the electronic properties of atoms, molecules and solids. It's success stems from the surprising accuracy of the earliest exchange-correlation (xc) approximation, the local-density approximation (LDA).\cite{KS1965} Going beyond LDA within the framework of the local density formalism leads to gradient-level density functionals (GDFs), which can be divided into two main families. The generalized gradient approximation\cite{WhyGGAworks} (GGA) managed to stabilize the diverging term in the second order gradient expansion\cite{KS1965} and gave e.g. a qualitatively correct description for the ground state of ferromagnetic iron.\cite{PRB-40-1997,PRB-43-11628,JPCM-2-7597} The subsystem functional approach (SFA)\cite{SFA} originates from the nearsightedness principle\cite{Nearsightedness} and incorporates inhomogeneous electron density effects through well-adapted model systems. For both GDF families, LDA represents the lowest order approximation and thus the correct limit in systems with densities showing negligible inhomogeneities. LDA together with the Perdew-Burke-Ernzerhof (PBE)\cite{PBE} GGA are the two most widely used xc-functionals for condensed matter.

In DFT, assuming proper numerical implementation of the software, the accuracy of the results for well converged calculations depends only on the chosen xc-functional. However, there is no functional that could provide systematic accuracy for a wide range of solids. In our previous work\cite{QNA} we have introduced the concept of quasi-non-uniform  gradient-level xc-approximation (QNA) and shown that it is able to produce highly accurate equilibrium volumes for metals and their alloys. This scheme is based on the observation that for metals the beyond-LDA features of a GGA-type xc-functional only matter in the valence-core overlap regions centered around the atomic sites. The error of a GDF is thus mainly local in nature. This finding concurs with the discussions in Refs. \onlinecite{PRB-57-2134,PRB-80-195109,PRB-79-155107} and it allows one to define a new SFA in which each (element specific) valence-core overlap region constitutes a subsystem and these subsystems are connected by the nearly homogeneous, LDA-like valence electron sea. For a general multicomponent system, then, one can apply local corrections in each core-valence overlap region separately and the overall QNA functional can mathematically be expressed as a superposition of the subsystem functionals, viz.
\begin{equation} \label{eq:QNA_equation}
E_{\text{xc}}^{\text{QNA}}[n]=\sum_q\int_{\omega_q}\epsilon_\text{x}^{\text{LDA}}(n)F_{\text{xc}}^{\text{opt}_q}(r_s,s)\,d^3r,
\end{equation}
where $r_s=[3/(4\pi n)]^{1/3}$, $s=|\nabla n|/[2n(3\pi^2n)^{1/3}]$ and $F_{\text{xc}}^{\text{opt}_q}$ in this case (see next paragraph) is the GGA enhancement function of the subsystem functional for the alloy component $q$. Around each atomic site $q$, the integration domain is within $\omega_q$. These space-filling polyhedra $\omega_q$ are defined so that the gradient of the density (and thus $s$) vanishes on the boundary of each $\omega_q$.

Defining the optimal subsystem functionals could be done in different ways. In this work we use perhaps the simplest procedure. Namely, all subsystem functionals have the same analytical form and only some ``tunable'' parameters are changed. Thus, for each element so called ``optimal'' parameters have to be found. In our previous work (see Ref. \onlinecite{QNA}) these optimal parameters were chosen in such a way that the error in equilibrium volume vanished. In this paper we extend our scope and investigate how well this process can be done when the error in equilibrium volume ($V_0$) and bulk modulus ($B_0$) are both taken into account. To this end, for each element we find such optimal parameters that minimize the objective function that we have chosen to be the combined absolute relative error:
\begin{equation} \label{eq:costfunction}
f(\text{*args})=W_a\frac{|a_0(\text{*args})-a_{\text{expt}}|}{a_{\text{expt}}}+W_B\frac{|B_0(\text{*args})-B_{\text{expt}}|}{B_{\text{expt}}},
\end{equation}
where $a$ is the lattice constant, $B$ is the bulk modulus, expt signifies experimental values and *args represents the tunable parameters that the functional form in use has. $W_a$ and $W_B$ are weights that can be chosen appropriately. We have chosen to use the framework of PBE and PBEsol\cite{PBEsol} due to its simple yet powerful construction. PBE was designed to provide accurate atomic energies whereas PBEsol was optimized for bulk and surface systems by restoring the original gradient expansion behavior for the exchange part and adjusting the correlation term using the jellium surface exchange-correlation energies obtained at meta-GGA level. Both PBE and PBEsol have two parameters, $\mu$ and $\beta$, which control the strength of exchange ($\mu$) and correlation ($\beta$) corrections over LDA. Their PBE and PBEsol values are $(\mu,\beta)_{\text{PBE}} =(0.219515,0.066725)$ and $(\mu,\beta)_{\text{PBEsol}}=(0.123457,0.046000)$, respectively. Finding optimal parameters, which is to say minimizing Eq. \eqref{eq:costfunction}, thus in this case becomes a two-dimensional optimization problem. In this work such optimization for 25 elements has been done. This group contains 10 cubic $sp$ metals and 15 transition metals. 

\section{\label{sec:method}Computational method}
The electronic structure and total energy calculations from which the optimal parameters were derived were performed employing the exact muffin-tin orbitals (EMTO) method.\cite{Andersen1994,VitosBook,EMTO2001PRB,*PRL-87-156401} Gradient corrections have been taken into account in a non-self-consistent (NSC) manner, which is to say self-consistent (SC) calculations were carried out within LDA and the gradient terms were included in the total energy within the perturbative approach.\cite{PerturbativeApproach} The reason behind favoring the NSC approach in this case is that it greatly reduces the required computational time. This way the resource intensive self-consistent Kohn-Sham loops can be completed first so that the optimization process only involves fast evaluations of the total energy. For Cr self-consistent GGA was employed instead of the perturbative approach. This is because for antiferromagnetic Cr the perturbative approach highly overestimates the bulk modulus and self-consistent GGA becomes necessary to correct for this error.\cite{PRB-65-184432} The Kohn-Sham equations were solved within the scalar relativistic approximation and the soft-core scheme. The Green’s function was calculated for 16 complex energy points distributed exponentially on a semicircular contour including the valence states and employing the double Taylor expansion approach.\cite{DoubleTaylor} The EMTO basis set included $s$, $p$, $d$, and $f$ orbitals $(l_{\textrm{max}}=3)$, and in the one-center expansion of the full charge density $l_{\textrm{max}}^h =8$ was used. The amount of inequivalent $\vec{k}$ points was $27434$ and $28884$ in the irreducible wedge of the body centered cubic (bcc) and face centered cubic (fcc) Brillouin zones, respectively. The theoretical equilibrium lattice constants $a_0$ and bulk moduli $B_0$ were derived from the stabilized jellium equations of state (SJEOS)\cite{SJEOS} fitted to the \emph{ab initio} total energies calculated for 18-21 atomic volumes around the equilibrium. For each element Eq. \eqref{eq:costfunction} was minimized with weights $W_a=W_B=1$ under the constraints of keeping the accuracy of the calculated Wigner-Seitz radius with respect to the experimental value roughly inside $\pm 0.005$ Bohr and not letting $\beta$ become greater than $0.1$.

To assess the transferability of these NSC-derived optimal parameters into SC-GGA calculations additional SC-EMTO and FP-(L)APW+lo Elk\cite{elk} calculations have been carried out for Li, V, Fe, Cu, Nb and Au. With Elk, a grid of $21\times21\times21$ $\vec{k}$ points (286 in the irreducible wedge) was used and $R_{\text{MT}}^{\text{min}}K_{\text{max}}$, which determines the size of the basis set, was between $8$ and $10$. Spin-orbit coupling has been taken into account for Au. Lattice constants and bulk moduli were obtained from a SJEOS-fit to $10$-$11$ points around the equilibrium.

Based on our experience the differences between NSC-EMTO and SC-EMTO results are expected to be quite small (see subsection \ref{subsec:optimal_params}). However, problems in the SC-EMTO results may occur if the numerical derivatives of the density (especially with the second derivative) are not being calculated with sufficient accuracy. Even small deficiencies in the way the derivatives are calculated can be detrimental, because the error has a tendency to amplify itself after each iteration of the self-consistent Kohn-Sham loop.
\begin{table*}
\caption{\label{tab:a0-B0-table}Theoretical and experimental equilibrium lattice constants $a_0$ ( in \AA) and bulk moduli $B_0$ ( in GPa) for the cubic $sp$, $3d$, $4d$ and $5d$ metals. Corresponding lattice structures are in parenthesis. The experimental data have been corrected for temperature and ZPPE terms. Results are shown for PBE, PBEsol and QNA functionals. The best theoretical values and statistical data are in boldface.}
\begin{ruledtabular}
\begin{tabular}{lcccccccc}
&\multicolumn{4}{c}{$a_0$}&\multicolumn{4}{c}{$B_0$}\\ \cline{2-5}\cline{6-9}
Solid&QNA&PBE&PBEsol&Expt.&QNA&PBE&PBEsol&Expt.\\ \hline
Li (bcc)&$\bm{3.444}$&$3.437$&$3.434$&$3.449$&$13.5$&$\bm{13.9}$&$13.7$&$13.8$\\
Na (bcc)&$\bm{4.204}$&$4.200$&$4.171$&$4.210$&$\bm{7.55}$&$7.77$&$7.91$&$7.63$\\
K  (bcc)&$\bm{5.212}$&$5.286$&$5.214$&$5.212$&$\bm{3.75}$&$3.58$&$\bm{3.74}$&$3.75$\\
Rb (bcc)&$\bm{5.581}$&$5.667$&$5.564$&$5.576$&$\bm{2.95}$&$2.82$&$2.99$&$2.92$\\
Cs (bcc)&$\bm{6.034}$&$6.167$&$6.016$&$6.039$&$\bm{2.07}$&$1.96$&$2.03$&$2.11$\\
Ca (fcc)&$\bm{5.546}$&$5.540$&$5.472$&$5.553$&$\bm{17.9}$&$17.3$&$\bm{17.9}$&$18.6$\\
Sr (fcc)&$\bm{6.038}$&$6.031$&$5.933$&$6.045$&$\bm{11.6}$&$\bm{11.6}$&$13.4$&$12.5$\\
Ba (bcc)&$\bm{4.991}$&$5.014$&$4.865$&$4.995$&$\bm{8.33}$&$8.09$&$8.28$&$9.34$\\
Al (fcc)&$\bm{4.018}$&$4.045$&$4.019$&$4.020$&$\bm{80.8}$&$76.8$&$81.3$&$80.8$\\
Pb (fcc)&$\bm{4.909}$&$5.053$&$4.947$&$4.902$&$48.3$&$39.3$&$\bm{46.5}$&$47.0$\\
V  (bcc)&$\bm{3.024}$&$2.998$&$2.958$&$3.024$&$\bm{163} $&$177 $&$190 $&$161 $\\
Cr (bcc)&$\bm{2.869}$&$\bm{2.869}$&$2.808$&$2.877$&$\bm{189}$&$184$&$252$&$194 $\\
Fe (bcc)&$\bm{2.854}$&$2.838$&$2.793$&$2.853$&$\bm{175} $&$189 $&$220$&$174 $\\
Ni (fcc)&$\bm{3.514}$&$3.526$&$3.470$&$3.508$&$200 $&$\bm{197} $&$228 $&$195 $\\
Cu (fcc)&$\bm{3.602}$&$3.637$&$3.571$&$3.595$&$150 $&$\bm{139} $&$167 $&$144 $\\
Nb (bcc)&$\bm{3.288}$&$3.310$&$3.269$&$3.294$&$167 $&$162 $&$\bm{169} $&$174 $\\
Mo (bcc)&$\bm{3.137}$&$3.164$&$3.131$&$3.141$&$\bm{266} $&$249 $&$\bm{266} $&$278 $\\
Rh (fcc)&$3.800$&$3.846$&$\bm{3.796}$&$3.793$&$\bm{283} $&$251 $&$288 $&$271 $\\
Pd (fcc)&$\bm{3.882}$&$3.958$&$3.890$&$3.875$&$204 $&$165 $&$\bm{201} $&$196 $\\
Ag (fcc)&$\bm{4.063}$&$4.163$&$4.068$&$4.056$&$119 $&$88.0$&$\bm{117} $&$110 $\\
Ta (bcc)&$\bm{3.295}$&$3.326$&$3.287$&$3.299$&$195 $&$187 $&$\bm{196} $&$198 $\\
W  (bcc)&$3.165$&$3.191$&$\bm{3.161}$&$3.160$&$\bm{305} $&$293 $&$311 $&$300 $\\
Ir (fcc)&$\bm{3.838}$&$3.890$&$3.850$&$3.831$&$386 $&$340 $&$\bm{377} $&$365 $\\
Pt (fcc)&$\bm{3.920}$&$3.988$&$3.936$&$3.913$&$\bm{295} $&$242 $&$282 $&$289 $\\
Au (fcc)&$\bm{4.068}$&$4.176$&$4.101$&$4.062$&$186 $&$135 $&$\bm{171} $&$178 $ \\ \hline
ME\footnote{Mean error.} (\AA $\times10^{-2}$)&$\bm{0.06}$&$4.15$&$-2.23$&\multicolumn{1}{l}{ME (GPa)}&$\bm{2.20}$&$-9.76$&$8.24$& \\
MAE\footnote{Mean absolute error.} (\AA $\times10^{-2}$)&$\bm{0.52}$&$4.95$&$3.52$&\multicolumn{1}{l}{MAE (GPa)}&$\bm{4.62}$&$12.38$&$11.06$& \\
MRE\footnote{Mean relative error.} (\%)&$\bm{0.02}$&$0.95$&$-0.56$&\multicolumn{1}{l}{MRE (\%)}&$\bm{0.14}$&$-6.67$&$4.28$& \\
MARE\footnote{Mean absolute relative error.} (\%)&$\bm{0.13}$&$1.17$&$0.87$&\multicolumn{1}{l}{MARE (\%)}&$\bm{3.32}$&$8.41$&$7.13$& \\
\end{tabular}
\end{ruledtabular}
\end{table*}

\section{\label{sec:results}Results and discussion}
All experimental lattice constants and bulk moduli have either been reported in 0 K or extrapolated to 0 K using the linear thermal expansion coefficients $\alpha$ from Ref. \onlinecite{CRC94}. Zero-point phonon effects (ZPPE) have also been subtracted out from both lattice constants and bulk moduli. 
For lattice constants, ZPPEs are in the form of zero-point anharmonic expansion (ZPAE) and it can be estimated as explained in Refs. \onlinecite{PRB-63-224115} and \onlinecite{PRB-69-075102} by using the expression
\begin{equation} \label{eq:ZPAE}
\frac{\Delta a_0}{a_0}=\frac{1}{3}\frac{\Delta V_0}{V_0}=\frac{3}{16}(B_1-1)\frac{k_B\Theta_D}{B_0V_{0,\text{at}}},
\end{equation}
where $\Delta V_0/V_0$ is the fractional volume change caused by the inclusion of ZPAE leading to a correction $\Delta a_0$ to the experimental lattice parameter $a_0$. $B_1$ is the pressure derivative of the bulk modulus $B_0$, $\Theta_D$ is the Debye temperature (from Ref. \onlinecite{Kittel}), and $V_{0,\text{at}}$ is the experimental volume per atom. With bulk moduli, the ZPPEs have been taken into account using the procedure of Ref. \onlinecite{GaudoinFoulkes}, according to which
\begin{equation} \label{eq:BmodZPPE}
\Delta B_0=B_1(P_t+P_z)=B_1\left(-\frac{\Delta V}{V}B-\frac{3}{16}B_1\frac{k_B\Theta_D}{B_0V_{0,\text{at}}}\right),
\end{equation}
where $P_t=-B\Delta V/V$ is a small negative pressure associated with the thermal expansion of a material and $P_z$ is the effective pressure required to mimic the effect of ZPPEs. In the present application, we used the data from Refs. \onlinecite{PRB-79-155107}, \onlinecite{PRB-79-085104} and the supplementary material from Ref. \onlinecite{PRB-83-205117} to estimate $B_1$ from Eq. \eqref{eq:ZPAE}. For Cr $B_1$ was estimated by using the data from Ref. \onlinecite{Gschneider}. The so derived $B_1$ values were used in Eq. \eqref{eq:BmodZPPE}. 

In this work a group of 25 metals is considered. This group contains monovalent $sp$ metals (Li, Na, K, Rb, and Cs), cubic divalent $sp$ metals (Ca, Sr, and Ba), Al, Pb and cubid $3d$ (V, Cr, Fe, Ni and Cu), $4d$ (Nb, Mo, Rh, Pd and Ag) and $5d$ (Ta, W, Ir, Pt and Au) metals. Experimental lattice constants are from Ref. \onlinecite{PRB-79-155107} (Li, Na, K, Rb, Cs, Ca, Sr, Ba, Al, Pb, Cu, Rh, Pd, and Ag), Ref. \onlinecite{PRB-79-085104} (V, Fe, Ni, Nb, Mo, Ta, W, Ir, Pt, and Au), and Ref. \onlinecite{CRC94} (Cr). Experimental bulk moduli are from Ref. \onlinecite{PRB-79-155107} (Li, Na, K, Rb, Cs, Ca, Sr, Ba, Al, Pb, Cu, Rh, Pd, and Ag), Ref. \onlinecite{PRB-3-4100} (V), Ref. \onlinecite{Kittel} (Cr, Ni, Nb, Mo, and Ir), Ref. \onlinecite{JoAP-100-113530} (Fe), and Ref. \onlinecite{PRB-70-094112} (Ta, W, Pt, and Au).

Table \ref{tab:a0-B0-table} contains the calculated equilibrium lattice constants and bulk moduli for the 25 elements considered as well as the 0 K estimated experimental values. Results for PBE, PBEsol and QNA functionals are included. Total mean error (ME), mean absolute error (MAE), mean relative error (MRE) and mean absolute relative error (MARE) for the lattice constants as well as the bulk moduli are also listed at the bottom of the table. 

\subsection{Lattice constants}
\begin{figure} 
\includegraphics[width=1.0\columnwidth]{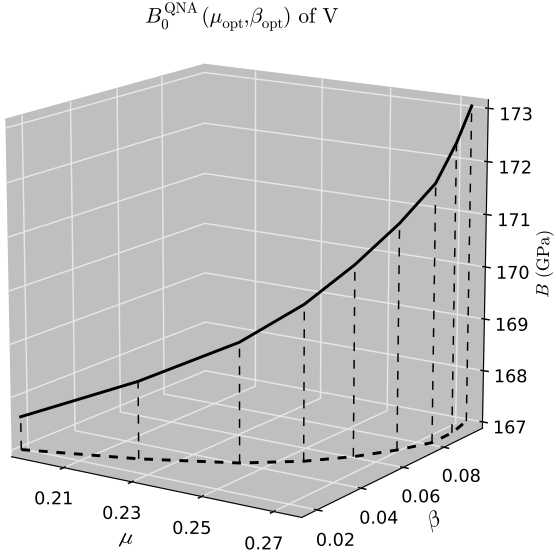}
\caption{QNA bulk modulus of V along the curve of such $\{\mu,\beta\}$ pairs, which all yield the same optimal volume (the thick dashed line).}
\label{fig:QNABmodV}
\end{figure}

The present trends for PBE and PBEsol are in line with investigations that have studied the performance of these functionals.\cite{AM05-applied,PRB-77-195445,
PRB-79-155107,PRB-79-085104,JoCTaC-9-1631,JoMM-19-2791} On average PBE tends to overestimate the volume while PBEsol does the opposite (see MEs and MREs in Table \ref{tab:a0-B0-table}). At least for the elements tested so far, it is always possible to find such optimal $\mu$ and $\beta$ that the error in lattice constant vanishes. In fact, there is an infinite amount of such $\{\mu,\beta\}$ pairs forming a continuous curve in the $\{\mu,\beta\}$ space (see Fig. \ref{fig:QNABmodV}). If the accuracy of the bulk modulus was not to be taken into consideration QNA's MAE in Table \ref{tab:a0-B0-table} for lattice constants would consequently be zero. Even with the accuracy of the bulk modulus factored in, MAE of QNA is an order of magnitude smaller than those of PBE and PBEsol. In all but two cases QNA is able to produce the most accurate lattice constant, with the exceptions being Rh and W. For Rh and W, PBEsol already gives a very accurate lattice constant and some of this accuracy has been given up in QNA to better match the error in bulk modulus. The effect in both cases, however, remains rather modest; QNA lattice constant is only $0.004$ Bohr larger that that of the best performing functional.

\begin{figure} 
\includegraphics[width=1.0\columnwidth]{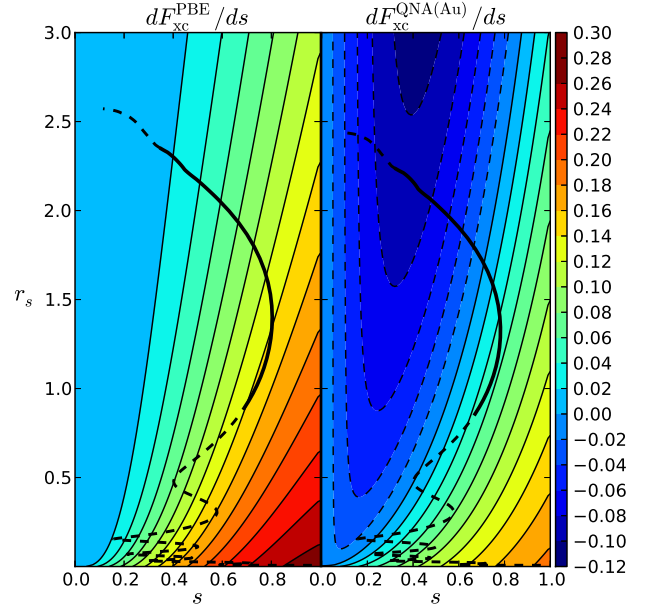}
\caption{\label{fig:dFcx_ds} (Color online) $s$-derivatives of the enhancement function $F_{\text{xc}}$ for PBE (left) and QNA(Au) (right). The thick black lines represent the values of $r_s$ and $s$ of Au in the $\hat{z}$-direction within the Wigner-Seitz cell at the PBE and QNA equilibrium volumes. The solid portion of these lines represent the core-valence overlap region. Dashed contour lines represent negative regions.}
\end{figure}

It has been explained in Ref. \onlinecite{PRB-80-195109} how the calculated equilibrium volume is determined by the slope $dE_{\text{xc}}/dV$, where $V$ is some measure for the volume. In this paper the Wigner-Seitz radius $w$ will be used. Different slopes for different functionals arise from the core-valence overlap region with larger slopes corresponding to smaller volumes and vice versa. By changing the parameters $\mu$ and $\beta$ one can manipulate the shape of the $F_{\text{xc}}^{\text{opt}_q}(r_s,s)$-map, which in turn is going to determine the slope $dE_{\text{xc}}/dw$. In the interesting region (core-valence overlap region) this slope is given by
\begin{equation} \label{eq:dExc_dOmega}
\frac{\partial E_{\text{xc}}^{\text{cvor}}}{\partial w}=A(G_1+G_2+G_3),
\end{equation}
where $A=(3/4)(3/\pi)^{1/3}[3/(4\pi)]^{4/3}$ and ``cvor'' stands for core-valence overlap region. Notation of Ref. \onlinecite{PRB-80-195109} has been used, in which
\begin{eqnarray}
G_1&=&\int_{\text{cvor}}\frac{4}{r_s^5}\frac{d r_s}{dw}F_{\text{xc}}\,d^3r,\label{eq:G1}\\
G_2&=&-\int_{\text{cvor}}\frac{1}{r_s^4}\frac{\partial F_{\text{xc}}}{\partial r_s}\frac{d r_s}{dw}\,d^3r,\label{eq:G2}\\
G_3&=&-\int_{\text{cvor}}\frac{1}{r_s^4}\frac{\partial F_{\text{xc}}}{\partial s}\frac{d s}{dw}\,d^3r.\label{eq:G3}
\end{eqnarray}
It is the interplay of $G_1$, $G_2$ and $G_3$ which determines the slope and as a result the lattice constant. $G_1$ is the strongest and positive, while $G_2$ and $G_3$ are generally one to two orders of magnitude weaker and negative in sign.\cite{PRB-80-195109} Note that for LDA $G_3$ is always zero since $\partial F_{\text{xc}}/\partial s$ is zero by definition. For PBE, on the other hand, the $\partial F_{\text{xc}}/\partial s$-term is always positive and relatively strong in the interesting region, as can be seen in Fig. \ref{fig:dFcx_ds}. This figure also displays the parametric curves of $r_s$ and $s$ of Au in the $\hat{z}$-direction within the Wigner-Seitz cell at the PBE and QNA equilibrium volumes. The solid portions of these curves represent the core-valence overlap region and they illustrate what kind of $\partial F_{\text{xc}}/\partial s$ values appear inside the integral of Eq \eqref{eq:G3} for PBE and QNA. As a result $G_3^{\text{PBE}}$ cancels a fair amount out of $G_1^{\text{PBE}}$ leading to shallow slopes and overestimated lattice constants for many solids. But Au, for example, has optimal parameters $\mu=0.125$ and $\beta=0.1$ leading to $\partial F_{\text{xc}}/\partial s$-map, which in the interesting region is weaker by roughly a factor of 2-3 compared to the $\partial F_{\text{xc}}/\partial s$-map of PBE (see Fig. \ref{fig:dFcx_ds}). It also has negative areas which partially cancel the contributions from positive areas even further undermining the significance of the $G_3^{\text{QNA(Au)}}$-contribution. This is why Au for instance prefers such a choice of optimal parameters, since they lead to an LDA-like steep $dE_{\text{xc}}/dw$ slope and thus a correct lattice constant. 

\subsection{Bulk moduli}

\begin{table*}
\caption{\label{tab:GH-table} Calculated values of $G_1$-$G_3$, $G_{\text{tot}}$, $H_1$-$H_6$ and $H_{\text{tot}}$ for V and Au using three different equivolume $\{\mu,\beta\}$-pairs.}
\begin{ruledtabular}
\begin{tabular}{lcccccc}
&\multicolumn{3}{c}{V, $a_0=3.024$ \AA}&\multicolumn{3}{c}{Au, $a_0=4.068$ \AA}\\ \cline{2-4}\cline{5-7}
$\mu$&$0.1880$&$0.2995$&$0.2945$&$0.0168$&$0.0969$&$0.1250$\\
$\beta$&$0.0050$&$0.0500$&$0.1000$&$0.005$&$0.0500$&$0.1000$\\
$B_0$ (GPa)&$163$&$167$&$170$&$188$&$188$&$187$\\\hline
$G_1$&0.856&0.854&0.840&0.614&0.608&0.597\\
$G_2$&-0.016&-0.015&-0.013&-0.012&-0.011&-0.009\\
$G_3$&-0.078&-0.077&-0.066&-0.001&-0.003&-0.006\\
$G_\text{tot}$&0.762&0.762&0.761&0.601&0.594&0.584\\\hline
$H_1$&-2.159&-2.155&-2.118&-2.004&-1.985&-1.953\\
$H_2$&0.052&0.047&0.041&0.046&0.040&0.032\\
$H_3$&0.183&0.187&0.169&0.003&0.009&0.018\\
$H_4$&0.000&0.005&0.008&0.000&0.006&0.012\\
$H_5$&0.003&0.002&0.002&0.001&0.001&0.001\\
$H_6$&-0.036&-0.037&-0.051&-0.001&-0.029&-0.068\\
$H_\text{tot}$&-1.957&-1.950&-1.948&-1.955&-1.958&-1.959
\end{tabular}
\end{ruledtabular}
\end{table*}

Unlike in Refs. \onlinecite{PRB-77-195445}, \onlinecite{PRB-79-085104}, \onlinecite{JoCTaC-9-1631}, and \onlinecite{JoMM-19-2791}, our results find the MAE (MUE in Ref. \onlinecite{JoMM-19-2791}) of PBEsol to be smaller than that of PBE. This is at least partly due to the use of ZPPE-corrected experimental bulk moduli. As PBEsol generally (except for Li in Table \ref{tab:a0-B0-table}) produces larger bulk moduli than PBE, ZPPE-corrections favour PBEsol, since these corrections increase the values of experimental bulk moduli. On average (MEs and MREs in Table \ref{tab:a0-B0-table}) PBE produces bulk moduli that are too small and PBEsol bulk moduli that are too large, while QNA has no strong bias towards either overestimation or underestimation. 

In terms of bulk modulus the improvements offered by QNA approximation are very clear. In most cases such good optimal parameters can be found that MAE and MARE of QNA in Table \ref{tab:a0-B0-table} are less than a half of the MAE and MARE of PBEsol. For many of the heavier elements, however, PBEsol produces the most accurate bulk modulus. Ba seems to be a difficult case as the relative error of the QNA bulk modulus remains at $11 \%$.

The value of the calculated bulk modulus of different functionals is mostly determined by the volume dependence of the total energy, which causes the error in the bulk modulus to be inversely related to the error in the lattice constant.\cite{PRB-76-024309} This is, for example, why  LDA tends to overestimate bulk moduli while PBE underestimates them. There is, however, a secondary effect which becomes important within QNA when the value of the bulk modulus is tuned while keeping the volume fixed to its optimal value. This secondary effect is the contribution coming from the curvature $\partial^2E_{\text{xc}}/\partial w^2$ of the $E_{\text{xc}}$ vs. $V$ curve. Increasing (decreasing) the negative curvature of the $E_{\text{xc}}$ vs. $V$ curve while keeping the overall slope (difference between the end points) fixed moves points higher (lower) in energy near the equilibrium volume and makes the $E_{\text{tot}}$ vs. $V$ curve shallower (deeper) thus giving smaller (higher) bulk modulus. The curvature arising from the interesting region is of the form
\begin{equation} \label{eq:d2Exc_dOmega2}
\frac{\partial^2 E_{\text{xc}}^{\text{cvor}}}{\partial w^2}=
A(H_1+H_2+H_3+H_4+H_5+H_6),
\end{equation}
where the six contributors $H_1$-$H_6$ have been grouped in terms of the enhancement factor $F_{\text{xc}}$ and it's derivatives and they are
\begin{eqnarray}
H_1&=&\int_{\text{cvor}}F_{\text{xc}}\left[\frac{4}{r_s^5}\frac{d^2r_s}{dw^2}-\frac{20}{r_s^6}\left(\frac{dr_s}{dw}\right)^2\right]\,d^3r,\label{eq:H1}\\
H_2&=&\int_{\text{cvor}}\frac{\partial F_{\text{xc}}}{\partial r_s}\left[\frac{8}{r_s^5}\left(\frac{dr_s}{dw}\right)^2-\frac{1}{r_s^4}\frac{d^2r_s}{dw^2}\right]\,d^3r,\label{eq:H2}\\
H_3&=&\int_{\text{cvor}}\frac{\partial F_{\text{xc}}}{\partial s}\left[\frac{8}{r_s^5}\frac{dr_s}{dw}\frac{ds}{dw}-\frac{1}{r_s^4}\frac{d^2s}{dw^2}\right]\,d^3r,\label{eq:H3}
\end{eqnarray}
\begin{eqnarray}
H_4&=&-\int_{\text{cvor}}\frac{\partial^2F_{\text{xc}}}{\partial s\partial r_s}\frac{2}{r_s^4}\frac{dr_s}{dw}\frac{ds}{dw}\,d^3r,\label{eq:H4}\\
H_5&=&-\int_{\text{cvor}}\frac{\partial^2F_{\text{xc}}}{\partial r_s^2}\frac{1}{r_s^4}\left(\frac{dr_s}{dw}\right)^2\,d^3r,\label{eq:H5}\\
H_6&=&-\int_{\text{cvor}}\frac{\partial^2F_{\text{xc}}}{\partial s^2}\frac{1}{r_s^4}\left(\frac{ds}{dw}\right)^2\,d^3r.\label{eq:H6}
\end{eqnarray}

Similarly to Eq. \eqref{eq:dExc_dOmega}, the biggest contribution comes from the negative $H_1$-term. Second most important terms are $H_2$, $H_3$ and $H_6$, while $H_4$ and $H_5$ most of the time yield practically negligible contributions, because the second partial derivatives of $F_{\text{xc}}$ involving $\partial/\partial r_s$ generally tend to  be small. $\partial F_{\text{xc}}/\partial s$- and $\partial^2F_{\text{xc}}/\partial s^2$-terms are much more sensitive to the details of the functional than their $r_s$ counterparts, making $H_3$- and $H_6$-contributions vary between high importance and insignificance depending on the actual case (see Table \ref{tab:GH-table}). $d^2r_s/dw^2$ and $d^2s/dw^2$ seem to be of the same order of magnitude with each other as well as with $dr_s/dw$ and $ds/dw$. For most elements tested the curvature becomes smaller in magnitude, which is to say the value of the bulk modulus increases, as we move higher in $\beta$ along the curve of fixed optimal volume in $\{\mu,\beta\}$-space. Fig. \ref{fig:QNABmodV} displays one such curve for V. The heaviest elements Pb and Au have the opposite behavior. Not as clear trend has been observed with elements having small bulk modulus, such as Li, which could be due to numerical difficulties associated with very shallow $E_{\text{tot}}$ vs. $V$ curves.

To better understand how these differences in curvatures as a function of optimal $\beta$ come about, we have approximated the terms $G_1$-$G_3$ and $H_1$-$H_6$ by calculating them in the $\hat{z}$-direction inside the core-valence overlap region within the Wigner-Seitz cell at the QNA equilibrium volume for V and Au. NSC-approach has been used, which means that $r_s$, $s$ and their first and second derivatives stay the same and only $F_{\text{xc}}$ and it's derivatives change between different sets of calculations.

\begin{table}
\caption{\label{tab:Optimal-parameters} Values of optimal parameters which minimize the combined error in lattice constant and bulk modulus (Eq. \eqref{eq:costfunction}) for the selected solids.}
\begin{ruledtabular}
\begin{tabular}{lcc}
Solid&$\mu_{\text{opt}}$&$\beta_{\text{opt}}$\\ \hline
Li (bcc)&$0.0878000$&$0.0718111$\\
Na (bcc)&$0.0960000$&$0.0000010$\\
K  (bcc)&$0.1187816$&$0.0472974$\\
Rb (bcc)&$0.1220000$&$0.0550631$\\
Cs (bcc)&$0.1333000$&$0.0180581$\\
Ca (fcc)&$0.1500000$&$0.1000000$\\
Sr (fcc)&$0.1470000$&$0.0050000$\\
Ba (bcc)&$0.1950000$&$0.0273746$\\
Al (fcc)&$0.1147214$&$0.0401048$\\
Pb (fcc)&$0.1260000$&$0.1000000$\\
V  (bcc)&$0.1880000$&$0.0050000$\\
Cr (bcc)&$0.0750000$&$0.0002000$\\
Fe (bcc)&$0.1485000$&$0.0050000$\\
Ni (fcc)&$0.1020000$&$0.0050000$\\
Cu (fcc)&$0.0795000$&$0.0050000$\\
Nb (bcc)&$0.1730000$&$0.1000000$\\
Mo (bcc)&$0.1600000$&$0.1000000$\\
Rh (fcc)&$0.0500000$&$0.0050000$\\
Pd (fcc)&$0.1415645$&$0.1000000$\\
Ag (fcc)&$0.1070000$&$0.0353330$\\
Ta (bcc)&$0.1450000$&$0.0654408$\\
W  (bcc)&$0.0605000$&$0.0050000$\\
Ir (fcc)&$0.0250000$&$0.0050000$\\
Pt (fcc)&$0.1130000$&$0.0561219$\\
Au (fcc)&$0.1250000$&$0.1000000$
\end{tabular}
\end{ruledtabular}
\end{table}

Table \ref{tab:GH-table} shows results for three different pairs of optimal $\mu$ and $\beta$ yielding the same volume as in Table \ref{tab:a0-B0-table}. 
The sums $G_\text{tot}=G_1+G_2+G_3$ and $H_\text{tot}=H_1+H_2+H_3+H_4+H_5+H_6$, respectively describing the slope and curvature of an $E_{\text{xc}}$ vs. $V$ curve, are in decent agreement with the observable trends relating to calculated volumes and bulk moduli as $\mu$ and $\beta$ are changed. $G_\text{tot}$ stays nearly constant (except for Au in the high-$\beta$ limit), as it should since all three pairs of $\mu$ and $\beta$ give identical equilibrium volumes. $G_\text{tot}$ also respects sequences $G_\text{tot}^{\text{LDA}}$$>$$G_\text{tot}^{\text{PBEsol}}$$>$$G_\text{tot}^{\text{PBE}}$$>$$G_\text{tot}^{\text{QNA(V)}}$ and $G_\text{tot}^{\text{LDA}}$$\approx$$G_\text{tot}^{\text{QNA(Au)}}$$>$$G_\text{tot}^{\text{PBEsol}}$$>$$G_\text{tot}^{\text{PBE}}$ (not shown), which agrees with the observed ordering of the lattice constants. 

For V (Au) $H_\text{tot}$ decreases (increases) in magnitude with increasing $\beta$, which agrees with the way bulk modulus evolves through increasing $\beta$. Despite $H_1$ being major part of $H_\text{tot}$, looking at $H_1$ alone is not enough to explain these trends. For Au all terms except $H_5$ have to be taken into account to make $H_\text{tot}$ slowly increase in magnitude as $\beta$ is increased.

\begin{table}
\caption{\label{tab:Elk_results} Self-consistent GGA lattice constants $a_0$ (in \AA) and bulk moduli (in GPa) calculated with EMTO and Elk using PBE and QNA functionals for Li, V, Fe, Cu, Nb, and Au. QNA is using the optimal parameters from Table \ref{tab:Optimal-parameters}. $\Delta_a$ (in \AA$\times10^{-3}$) and $\Delta_B$ (in GPa) indicate the difference compared to the NSC-EMTO results of Table \ref{tab:a0-B0-table} for QNA (``$\text{NSC}-\text{SC}$'' or ``$\text{NSC}-\text{Elk}$'').}
\begin{ruledtabular}
\begin{tabular}{lcccccc}
\multicolumn{7}{c}{EMTO}\\
&\multicolumn{3}{c}{$a_0$}&\multicolumn{3}{c}{$B_0$}\\ \cline{2-4}\cline{5-7}
Solid&QNA&PBE&$\Delta_a$&QNA&PBE&$\Delta_B$\\ \hline
Li (bcc)&$3.442$&$3.433$&$2$&$13.5$ &$14.0$&$ 0$\\
V (bcc) &$3.023$&$2.996$&$1$&$164  $ &$176 $&$-1$\\
Fe (bcc)&$2.854$&$2.836$&$0$&$179  $ &$193 $&$-4$\\
Cu (fcc)&$3.602$&$3.637$&$0$&$150  $ &$139 $&$ 0$\\
Nb (bcc)&$3.286$&$3.308$&$2$&$165  $ &$160 $&$ 2$\\
Au (fcc)&$4.069$&$4.177$&$-1$&$185 $ &$135 $&$ 1$\\
\multicolumn{7}{c}{}\\
\multicolumn{7}{c}{Elk}\\
&\multicolumn{3}{c}{$a_0$}&\multicolumn{3}{c}{$B_0$}\\ \cline{2-4}\cline{5-7}
Solid&QNA&PBE&$\Delta_a$&QNA&PBE&$\Delta_B$\\ \hline
Li (bcc)&$3.442$&$3.433$&$ 2$&$13.5$&$13.9$&$0$\\
V  (bcc)&$3.027$&$3.002$&$-3$&$170$&$181$&$-7$\\
Fe (bcc)&$2.850$&$2.834$&$ 4$&$177$&$194$&$-2$\\
Cu (fcc)&$3.598$&$3.632$&$ 4$&$153$&$141$&$-3$\\ 
Nb (bcc)&$3.294$&$3.314$&$-6$&$176$&$169$&$-9$\\
Au (fcc)&$4.048$&$4.149$&&$199$&$145$&\\\end{tabular}
\end{ruledtabular}
\end{table}

\subsection{\label{subsec:optimal_params}Optimal parameters}
\begin{figure} 
\includegraphics[width=1.0\columnwidth]{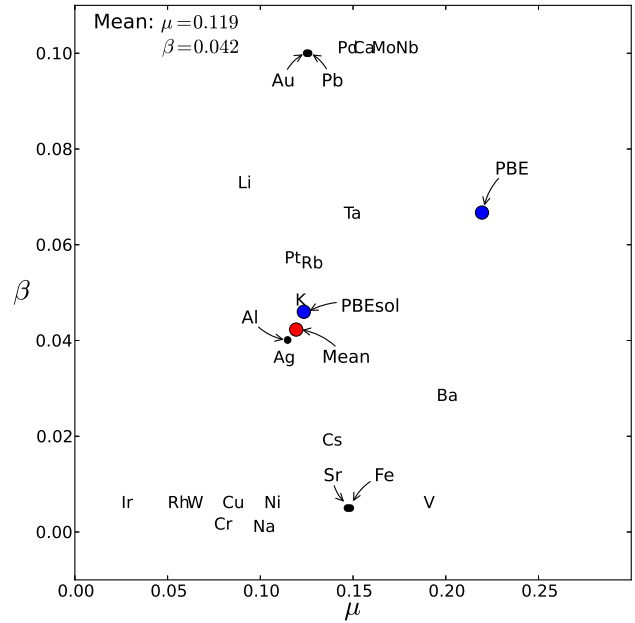}
\caption{\label{fig:Optimal_params_map} (Color online) Optimal parameters for the selected monoatomic solids. PBE and PBEsol points as well as the mean of the QNA optimal parameters are also drawn as annotated.}
\end{figure}

Optimal parameters minimizing the combined error in lattice constant and bulk modulus for the selected elements are presented in Table \ref{tab:Optimal-parameters}. They are also laid out graphically in $\{\mu,\beta\}$-space in Fig. \ref{fig:Optimal_params_map}. Best optimal values yielding nearly vanishing errors can generally be obtained in cases where either PBE or PBEsol tends to underestimate the lattice constant but overestimate bulk modulus, or vice versa. This feature enables efficient matching of the volume and bulk modulus errors, since increasing volume generally decreases the value of bulk modulus and vice versa. Note that while it is possible to completely minimize the error either in lattice constant or bulk modulus, it is much more difficult to completely minimize both errors at the same time. The presently employed PBE/PBEsol functional form is not flexible enough to allow for that.

In light of the mean errors in Table \ref{tab:a0-B0-table} constructing a functional out of element specific subfunctionals is one clear way of improving accuracy, while it has proven to be difficult to design an element independent GGA-level functional form that would have a consistent performance across the periodic table.\cite{PRB-79-085104,QNA} The difficulty lies in the fact that while some pair of elements assume fairly similar values of $r_s$ and $s$ in the core-valence overlap region, they might require very differently shaped $F_{\text{xc}}^{\text{opt}_q}(r_s,s)$-maps. As a result an element independent functional with only one $F_{\text{xc}}(r_s,s)$-map would have to be able to change it's shape very rapidly and non-trivially as a function of $r_s$ and $s$ over relatively short distances.

For the sake of reducing computational time the optimization process was done using NSC-GGA, i.e., self-consistent calculations were performed at LDA level and gradient corrections were included in the total energies as perturbations. To test the validity of these NSC-derived optimal parameters in conjunction with a fully self-consistent method further SC-EMTO and FP-(L)APW+lo Elk calculations for Li, V, Fe, Cu, Nb and Au were carried out. Results of these calculations along with the accompanying experimental values are presented in Table \ref{tab:Elk_results}. 

The differences between NSC-EMTO and SC-EMTO lattice parameters are very small. Bulk moduli show minor deviations with the $\sim-4$ GPa difference of Fe being the most notable. Similar observation concerning the discrepancy between the NSC and SC bulk moduli of Fe has been made in Ref. \onlinecite{PerturbativeApproach} where it was attributed to a connection between the gradient effects and the nonspherical spin densities of Fe. In our case, however, the SC treatment increases the bulk modulus of Fe, while in Ref. \onlinecite{PerturbativeApproach} (all-electron FP-KKR) it decreased.

Elk results in general agree very well with the NSC-EMTO calculations and thus with the experimental values. For Au the Elk lattice parameter becomes too small and bulk modulus too high, but this is mostly due to the inclusion of spin-orbit coupling, which was not present in the NSC-EMTO calculations. Elk bulk moduli show some differences. This is expected, since by comparing, for example, the results of Table \ref{tab:a0-B0-table} and Ref. \onlinecite{PRB-77-195445} (PBE, EMTO) to the results in the supplementary material of Ref. \onlinecite{PRB-83-205117} (PBE, WIEN2K) one can see that bulk moduli in Elk and in methods similar to Elk (FP-(L)APW+lo) tend to be higher than their EMTO counterparts for many of the $3d$, $4d$ and $5d$ metals. For solids such as Nb, for which QNA underestimates the bulk modulus, this is only beneficial. On the other hand, for solids with overestimated NSC-QNA bulk moduli the SC-QNA bulk modulus can be further overestimated depending on the employed computational method. Therefore, depending on the solid and the used computational method, it might be necessary to further optimize the values given in Table \ref{tab:Optimal-parameters} to make them better suited for SC-QNA calculations. For example, optimal $\mu$ for Au should be increased from its value of $0.125$ to compensate for the inclusion of spin-orbit coupling. Generally speaking, increasing $\mu$ will increase (decrease) the calculated lattice constant (bulk modulus) and vice versa, while increasing $\beta$ tends to decrease (increase) the calculated lattice constant (bulk modulus) up to some element specific point, after which the trend is reversed (see Fig. 2 of Ref. \onlinecite{QNA}).

\begin{table}
\caption{\label{tab:Alloys} Lattice constants $a_0$ (in \AA) and bulk moduli (in GPa)  for bcc $\text{Fe}_{1-x}\text{V}_x$ and B2 NiAl calculated with EMTO using PBE and QNA functionals. QNA is using the optimal parameters from Table \ref{tab:Optimal-parameters}. Experimental values from Ref. \onlinecite{PRB-82-144306} (FeV, room temp.) and Ref. \onlinecite{PRB-79-085104} (NiAl, 0 K) are also included.}
\begin{ruledtabular}
\begin{tabular}{lcccccc}
\multicolumn{7}{c}{EMTO}\\
&\multicolumn{3}{c}{$a_0$}&\multicolumn{3}{c}{$B_0$}\\ \cline{2-4}\cline{5-7}
Solid&QNA&PBE&Expt.&QNA&PBE&Expt.\\ \hline
NiAl (B2)&$2.881$&$2.894$&$2.882$&$163$&$159$&$156$\\
\multicolumn{7}{c}{}\\
Fe (bcc)&$2.854$&$2.838$&$2.858$&$175$&$189$&$170$\\
$\text{Fe}_{75}\text{V}_{25}$ (bcc)&$2.889$&$2.870$&$2.886$&$171$&$184$&$166$\\
$\text{Fe}_{50}\text{V}_{50}$ (bcc)&$2.919$&$2.897$&$2.914$&$169$&$183$&$161$\\
$\text{Fe}_{25}\text{V}_{75}$ (bcc)&$2.948$&$2.924$&$2.963$&$188$&$202$&$182$\\
$\text{Fe}_{6}\text{V}_{94}$ (bcc) &$3.004$&$2.978$&$3.020$&$169$&$183$&$172$\\
V (bcc)&$3.024$&$2.998$&$3.042$&$163$&$177$&$163$\\
\end{tabular}
\end{ruledtabular}
\end{table} 

It is interesting to note how in Fig. \ref{fig:Optimal_params_map} the mean value of the optimal parameters nearly coincides with the PBEsol parameters. If the restriction $\beta\leq0.1$ were to be lifted to allow Ca, Nb, Mo, Pd, Au, and Pb (see Fig. \ref{fig:Optimal_params_map} and Table \ref{tab:Optimal-parameters})  to have even higher values of optimal $\beta$, the agreement between the mean and PBEsol parameters would be even more convincing. It is not, however, surprising that the mean should be closer to PBEsol parameters than those of PBE, since PBEsol was designed for solids, so perhaps the mean of the optimal parameters can thus be viewed as an alternative confirmation of the PBEsol parameters.

\subsection{\label{sec:applications}Applications}
The performance of QNA scheme is tested on random $\text{Fe}_{1-x}\text{V}_x$ solid solution and ordered NiAl intermetallic compound having bcc and B2 structures, respectively. Previously, good performance for VW solid solution and CuAu (L$1_0$)  and Cu$_3$Au (L$1_2$) intermetallic compounds has been reported.\cite{QNA} Here FeV was treated using the NSC-GGA approach and coherent potential approximation\cite{CPA1,CPA2} and for both FeV and NiAl optimal parameters from Table \ref{tab:Optimal-parameters} were used. The experimental values are from Ref. \onlinecite{PRB-82-144306} (FeV) and Ref. \onlinecite{PRB-79-085104} (NiAl). For FeV the experimental values have not been corrected to 0 K (room temperature). The experimental values of NiAl are extrapolated to 0 K and the lattice constant includes ZPAEs but the bulk modulus does not include ZPPEs. The results are gathered in Table \ref{tab:Alloys}. It is no surprise that PBE underestimates the lattice constant of FeV, since it does so for both of the constituents. This results in the PBE bulk modulus being too large, as well. QNA corrects for these errors, improving the description of both quantities. At high V concentrations QNA lattice constant starts to deviate from the experimental value, which is due to the fact that the optimal parameters of V were calculated using the 0 K ZPAE-corrected experimental value ($3.024$ \AA) which is markedly smaller than the room temperature value of Table \ref{tab:Alloys} ($3.042$ \AA). 0 K corrections to the experimental values would significantly improve the accuracy at the V rich end. 

PBE lattice constant of NiAl is too large but bulk modulus is reproduced quite accurately due to cancellation of errors. QNA on the other hand gives an accurate lattice constant at the expence of slightly overestimated bulk modulus. We would like to highlight that the optimal parameters of Al and Ni were determined for their equilibrium fcc structures whereas NiAl adopts a bcc-like B2 structure. Hence, our findings demonstrate that at ambient conditions the optimal parameters are not dependent on the crystal structure and chemical environment.  However, they might be sensitive to extremely high pressure, which is a question to be investigated in the future.

\section{\label{sec:conclusions}conclusions}
We have investigated the flexibility of the quasi-non-uniform xc-approximation and
shown it to be able to significantly improve the description of lattice constants and bulk moduli over the currently used popular GGA-functionals for a large set of metals and their alloys. QNA achieves this by applying local corrections separately in each core-valence overlap region of the system at hand. Designing a uniform, element independent functional that would have a similar, consistent accuracy across the periodic table seems to be a rather difficult task. For any element tested, it is possible to completely minimize the error either in calculated volume or the bulk modulus but it is much more difficult to completely minimize both errors at the same time. The presently employed PBE functional form is not flexible enough to allow this. 

\begin{acknowledgments}
H. L. thanks Matti Ropo for providing comments and feedback. L. V. acknowledges the financial support from Swedish Research Council, the Swedish Steel Producers’ Association, the European Research Council, and the Hungarian Scientific Research Fund (research project OTKA 84078 and 109570).
The computer resources of the
Finnish IT Center for Science (CSC) and the FGI project
(Finland) are acknowledged.
\end{acknowledgments}

\end{document}